# Seeing Black Holes :
# from the Computer to the Telescope

Jean-Pierre Luminet
*Aix-Marseille Université, CNRS, Laboratoire d'Astrophysique de Marseille (LAM) UMR 7326*
*& Centre de Physique Théorique de Marseille (CPT) UMR 7332*
*& Observatoire de Paris (LUTH) UMR 8102*
*France*
*E-mail:* jean-pierre.luminet@lam.fr

**Abstract.** Astronomical observations are about to deliver the very first telescopic image of the massive black hole lurking at the Galactic Center. The mass of data collected in one night by the Event Horizon Telescope network, exceeding everything that has ever been done in any scientific field, should provide a recomposed image during 2018. All this, forty years after the first numerical simulations done by the present author.

## 1. Introduction:

According to the laws of general relativity, black holes are, by definition, invisible. Contrary to uncollapsed stars, their surface is neither a solid nor a gas; it is an intangible frontier known as the event horizon. Beyond this horizon, gravity is so strong that nothing escapes, not even light. Seen projected onto the background of the sky, the event horizon will probably resemble a perfectly black disk if the black hole is static (Schwarzschild black hole) or a slightly flattened disk if it is rotating (Kerr black hole).

A black hole however, be it small and of stellar mass or giant and supermassive, is rarely "bare"; in typical astrophysical conditions, it is usually surrounded by gaseous matter. This matter forms an accretion disk in which highly heated gas emits a characteristic radiation. A giant black hole, as can be found in the center of most galaxies, may also be surrounded by a cluster of stars, the orbital dynamics of which is strongly influenced by it. In essence, a black hole remains invisible, but in its own special way, it lights up the matter it attracts.

Logically, scientists wondered what a black hole lit up by its surrounding matter would look like. Educational or artistic representations can be seen in popular science magazines in the form of a sphere seeming to float in a whirlpool of glowing gas. These images, however forceful, fail to convey the astrophysical reality. It can be described correctly using computer simulations that take account of the complex distortions made by the gravitational field on space-time and on the paths of light rays that follow its fabric. These were performed by the first time in 1978 by the author of this article. Today, progress in astronomical observation is about to deliver the first telescopic image of a giant black hole, thanks to the ambitious Event Horizon Telescope programme.

## 2. Black holes simulated

To create the most realistic possible images of a black hole lit up by an accretion disk, not only do we have to calculate the propagation of light rays in the curved space-time geometry generated by the black hole, but we also have to know the physical properties of the accretion disk. In 1978, I was a young scientist at the Paris-Meudon Observatory and performed the first accurate numerical simulation of the "photographic" appearance of a Schwarzschild black hole surrounded by a thin accretion disk. To do so, I used the IBM 7040 mainframe of Paris-Meudon Observatory, an early transistor computer with punch card inputs. Without a computer visualisation tool, I had to create the final image by hand from the digital data. For this I drew directly on negative image paper with black India ink, placing dots more densely where the simulation showed more light. Next, I took the negative of my negative to get the positive, the black points becoming white and the white background becoming black.

This image (figure 1) appeared first in the November issue of a French popular magazine [1] and concluded a 1979 article in a specialized journal, with all equations and technical details [2].

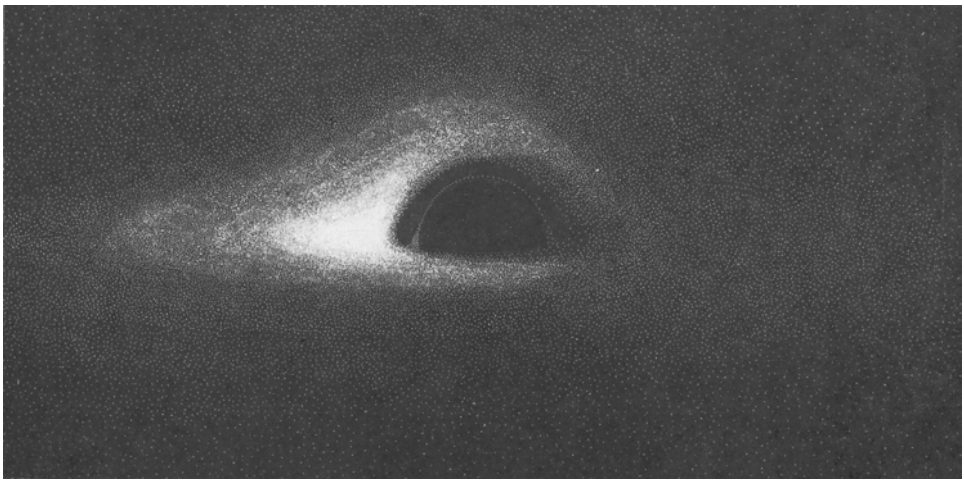

**Figure 1**. Simulated photograph of a spherical black hole with thin accretion disk. The system is seen from a great distance by an observer at 10° above the disk's plane.

The top of the disk remains visible regardless of the viewing angle - in contrast to the typical views of Saturn's rings. Indeed the gravitational field curves the light rays near the black hole so much that the rear part of the disk is "revealed". Even if the black hole hides what falls into it, it cannot mask what is behind it.

The curving of the light rays also generates a secondary image which allows us to see the other side of the accretion disk, on the opposing side of the black hole from the observer. Very deformed optically, the rear part looks like a thin halo of light against the central black disk, which represents the event horizon enlarged by a factor of $3\sqrt{3}/2 \approx 2{,}6$ due to the gravitational lens effect.

The main feature of this view of the black hole is the significant difference in luminosity between the various regions of the disk. On the one hand, the light shines maximum in the areas closest to the horizon, as the gas is hottest there. On the other hand, the light received by a distant observer is considerably different from the light emitted, due to the combination of the Einstein and Doppler effects; the first caused by the gravity field, the latter by the rapid rotation of the accretion disk. For a distant observer, the light received is considerably amplified on the side of the image where the gas approaches the observer and is weaker on the side it is moving away from.

The virtual photo of the black hole, calculated "bolometrically" (i.e. taking account of the whole electromagnetic spectrum from the radio to the gamma wave range) is independent of the mass of the black hole and the flow of gas swallowed, on the condition that this remains moderate (in other situations, the structure of the accretion disk may be thick, take the form of a torus, etc.). This image may also therefore describe a stellar black hole 10km in radius, attracting the gas from an accompanying star, or a giant black hole lying at the centre of a galaxy and sucking in the interstellar gas.

This initial digital imaging work on black holes was then developed by numerous scientists, who benefitted from rapid progress in computer performances. Colours were added to the images (according to a specific coding dictated by variations in temperature) and background skies, to make the reconstitutions as realistic as possible. Moreover, the observer is no longer supposedly to be stationary and very distant from the black hole, but moving with it, which introduces a new distortion of the images by the Doppler effect due to the movement of the observer. Finally, the black hole can be animated by a rotating movement, such as in the Kerr solution, the most realistic astrophysical situation. However, although this rotation generates an additional asymmetry - the black hole is no longer spherical - it remains small, even if rotating rapidly.

Of the numerous visualisations created, some of which are visible on the internet [3], those made at the start of the 1990s by my colleague Jean-Alain Marck at the Paris-Meudon Observatory, in colour and animated, are the most remarkable (for a technical description, [4]). We made a film which shows the spectacle that would be seen from the window of a spacecraft falling freely towards the black hole on various trajectories [5].

In the autumn of 2014, the world's media waxed lyrical on the representations of the supermassive black hole, "Gargantua", imagined by the film maker, Christopher Nolan, for his film *Interstellar.* The American astrophysicist Kip Thorne, a renowned specialist in relativistic astrophysics (he received the Nobel Prize for physics in 2017 for his work on gravitational waves), was technical advisor to a team of 200 graphic animation experts, using the most sophisticated calculation and visualisation tools, to model the appearance of a giant black hole measuring one hundred million solar masses, rotating rapidly and surrounded by an accretion disk.

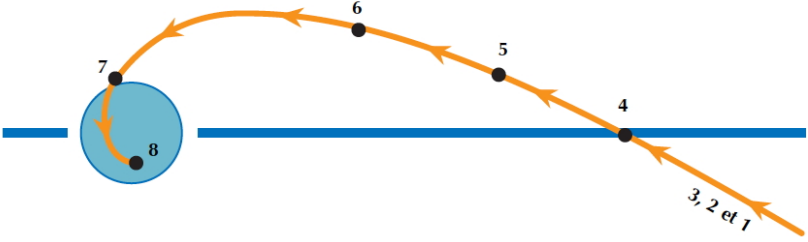

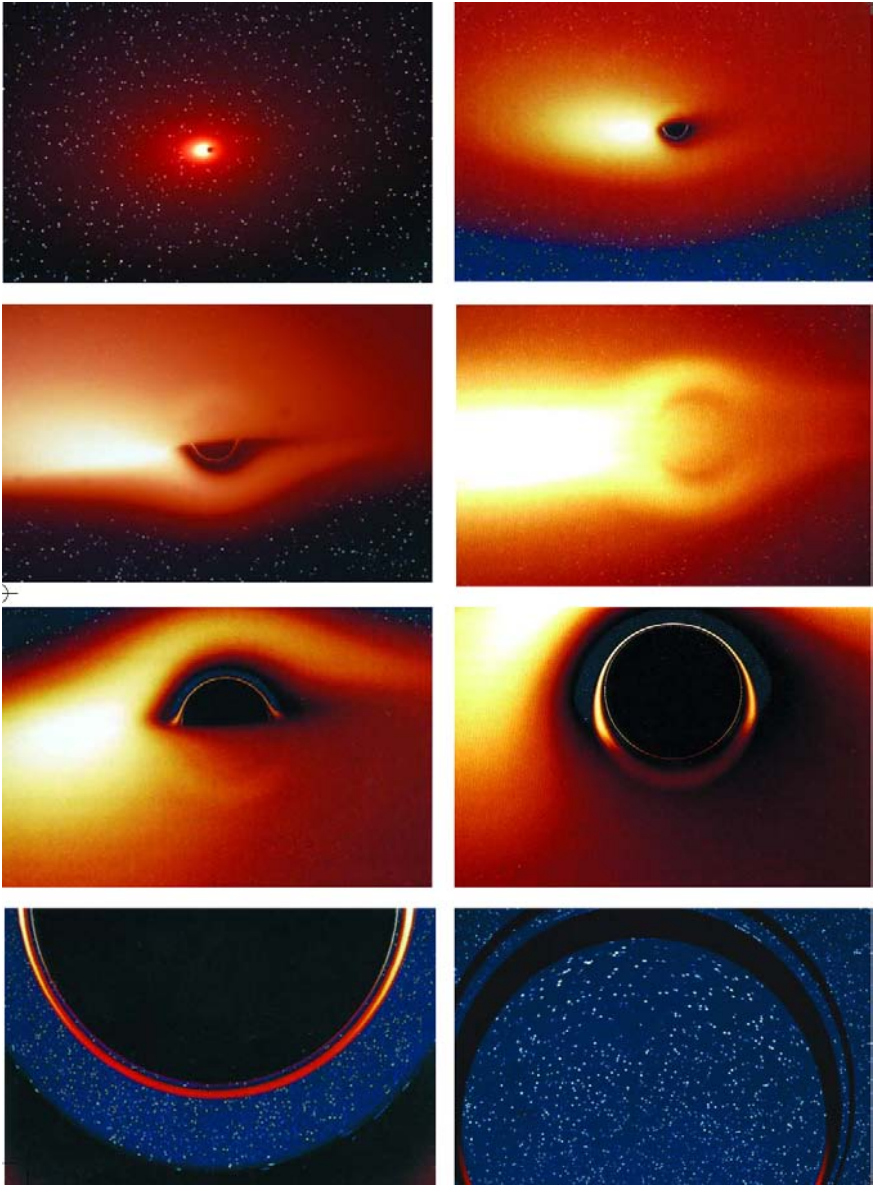

**Figure 2.** Colour simulations of a black hole accretion disk taking account of the Doppler and gravitational shifts. The images are calculated at positions 1 to 8 of the plunging trajectory schematized above. The last picture is taken from inside the black hole, the observer having rotated by 180° to watch the outside [6].

The international press headlined the scientific realism of the calculated images, resulting from a "*simulation of unprecedented accuracy*". The most captivating view in the film, calculated for an observer located in the plane of the accretion disk (figure 3), while it correctly describes the primary and secondary images distorted by the gravitational field, shows uniform luminosity of the disk.

In other words, it neglects all the physical effects due to its radiating structure and its rapid rotation. As Kip Thorne explained to me in a private email, the film maker decided that light asymmetry in the image would have been incomprehensible to the spectators! However, it is precisely this strong asymmetry of apparent luminosity that is the main signature of a black hole, the only celestial object able to give the internal regions of an accretion disk a speed of rotation close to the speed of light and to induce a very strong Doppler effect!

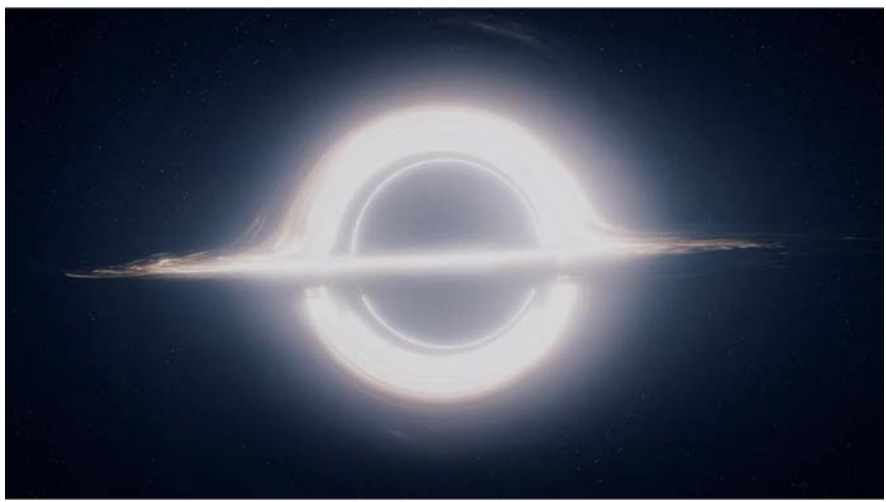

**Figure 3**. Simulation of an accretion disk around a Kerr black hole as seen by an observer in the equatorial plane, shown in the movie *Interstellar*. Only isoradial curves have been depicted, so that this is not a realistic image. Crédit: Double Negative artists/DNGR/TM&© Warner Bros.

Slightly embarrassed by this bending of the scientific truth, Thorne and his colleagues subsequently published an image in a technical journal, taking account of the effects of the spectral shifts (figure 4), but still based on a non-physical model of the accretion disk [7].

Indeed, from the computer simulations by Marck performed 20 years earlier, image 4 from figure 2 would be much closer from astrophysical reality, as it was also calculated seen by an observer in the equatorial plane, with all the shift effects AND a truly physical model of the accretion disk. For a more detailed criticism of *Interstellar* science, see [8].

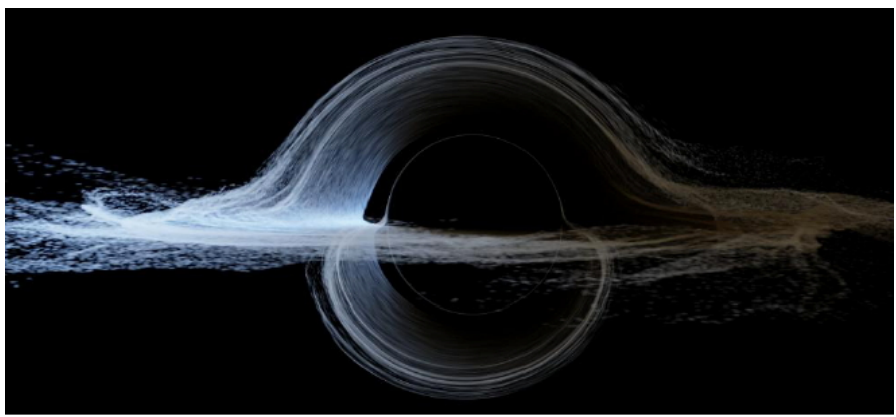

**Figure 4**. An « anemic » accretion disk around a Kerr black hole with spin a/M = 0.6, false colours, Doppler and gravitational shifts (from [6]).

Another way of visualising a "bare" black hole (without an accretion disk), is to calculate the gravitational mirage it causes on the background of stars. The most spectacular views, combining scientific accuracy and aesthetics, were obtained in 2006 by Alain Riazuelo [9]. He calculated the gravitational mirage caused by a black hole passing in front of a background of stars, the disk of our own galaxy or Magellanic clouds.

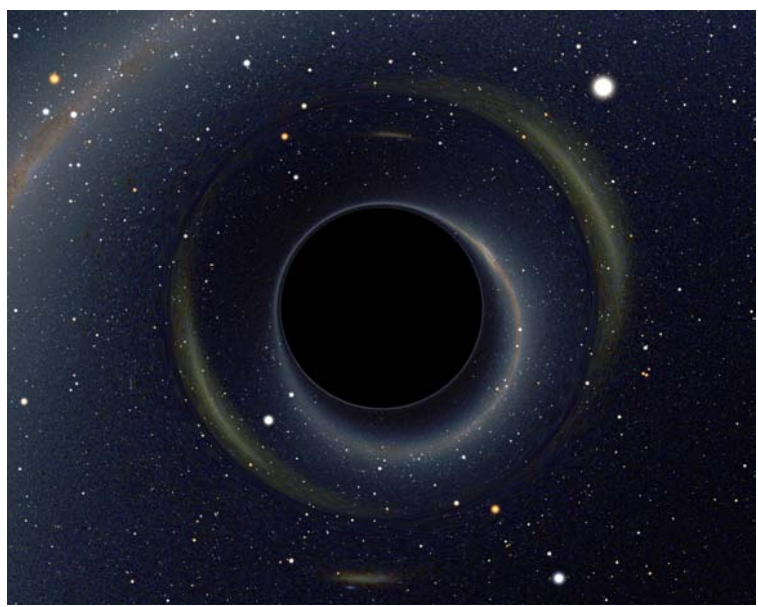

**Figure 5.** Gravitational lensing produced by a black hole in a direction almost centered on the Large Magellanic Cloud. Above it one easily notices the southernmost part of the Milky Way with, from left to right, Alpha and Beta Centauri, the Southern Cross. The brightest star close to the LMC is Canopus (seen twice). The second brightest star is Achernar, also seen twice. © Alain Riazuelo, CNRS/IAP

**From the computer to the telescope**

All these illustrations are virtual images obtained using the equations of general relativity and on more or less realistic physical models. But could we see a black hole directly? If astronomers had a sufficiently powerful telescope, they would be able to directly observe the shadow cast by the event horizon of a black hole and the hot mark of the accretion disk surrounding it. However, several technical challenges prevent the development of such an instrument. The main one is the tiny size of such stars seen from the Earth. The closest known stellar black hole, located in the binary X-ray source A0620-00 in our Galaxy, 3,500 light-years away, has a diameter of just 40 kilometres! So, we have to aim at close, supermassive black holes, knowing that their size is proportional to their mass. The two most promising candidates are Sagittarius A* (abbreviated to Sgr A*), located 26,000 light-years away in the centre of our galaxy, with a mass estimated at 4.3 million solar masses [10], and the supergiant M87*, a monster of 6 or 7 billion solar masses, lying at the centre of the giant elliptical galaxy of M87, 55 million light-years away [11]. Everything points to the fact that these black holes are surrounded by a rapidly-rotating accretion disk, possibly formed of star debris previously broken up by tidal forces.

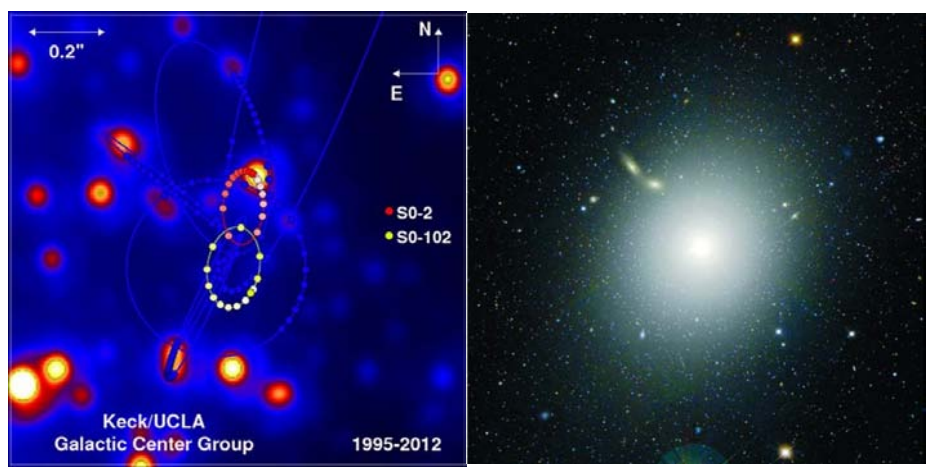

**Figure 6. The Galactic Center (left) and the giant elliptical galaxy M87 (right)**
Left : The trajectories of several stars orbiting around the Galactic Center have been plotted from infrared observations continuously performed from 1995 to 2012 by the Keck telescopes in Hawaii. Their dynamical analysis implies the existence of a central massive black hole, SgrA*, about 4 million solar masses. Right : The giant elliptical galaxy Messier 87, located in the Virgo local supercluster, shows at its very center a peak of luminosity, interpreted as the intense emission of gaz falling into supermassive black hole about 6 billion solar masses.

In terms of intrinsic size, the event horizons of Sgr A* and M87* are respectively 25 million and 36 billion kilometres in diameter. However, the effect of the gravitational lens caused by the black hole amplifies the apparent size of the event horizon by a factor of 2.6. All calculations done, it appears that the shadows cast by Sgr A* and M87* on the gaseous halos of their accretion disks have an apparent diameter of about 50 microarcseconds. This is the angle under which we would see an apple on the Moon with the naked eye, requiring a resolution 2,000 times greater than that of the Hubble space telescope!

The resolution of a telescope depends on its aperture (the diameter of its lens) and the wavelength at which it is observing. Doubling the aperture would show details twice as accurately. We would thus need a telescope operating in the visible range 2 kilometres in diameter to resolve the images of Sgr A* and M87*;

not possible, even in the medium-term. Another difficulty, is that the neighborhoods of black holes remain hidden from our view in most frequency bands of the electromagnetic spectrum. The galaxy centres where they lie are buried under dense clouds of dust that block out most of the radiation. To pierce through this fog, the wavelength needs to be increased.

Thus, at the millimetre wavelengths typical in radioastronomy, galaxy centres become almost transparent. Problem: when the wavelength is doubled, resolution is divided by two, such that the size of the telescopes needs to be increased further. So, to be able to observe the central black hole in our galaxy in the millimetre domain, we would need a radiotelescope about 5,000 km in diameter... Impossible? Not at all, because at these wavelengths, astronomers can use very large base interferometry (VLBI), a technique that combines several observatories into a single virtual telescope with an aperture as large as the distance that separates them. A terrestrial-sized VLBI network is therefore just sensitive enough to resolve an image as small as 50 microseconds of arc angle. This is how the EHT (Event Horizon Telescope) project was designed in the first decade of this century, combining millimetric radiotelescopes across the planet in a network, in the hope of capturing the first "real" images of giant black holes.

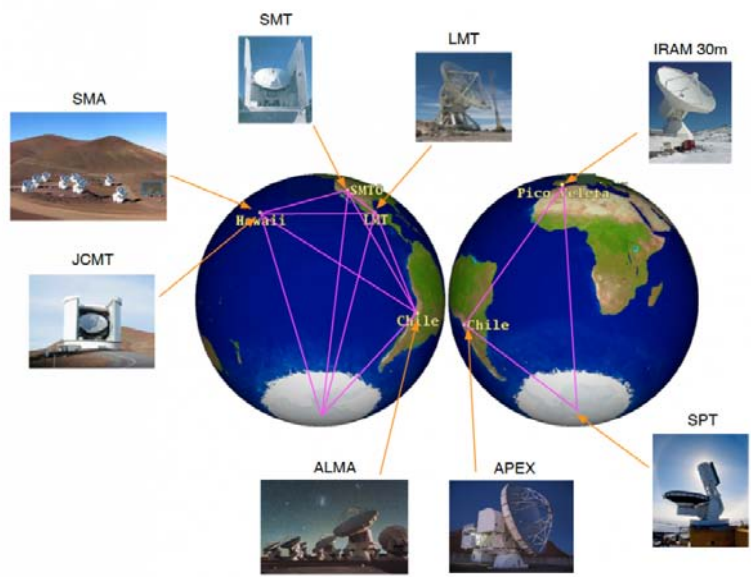

**Figure 7.** The VLBI network of the Event Horizon Telescope
(courtesy EHT team)

The idea of creating a worldwide network to observe the black hole at the centre of the galaxy started in 1999 at the initiative of the Dutch astronomer Heino Falcke [12], today at Nimègue University in the Netherlands. It was further developed by various radioastronomy groups [13], which merged in 2006 into the EHT consortium [14]. Gradually, the network grew to include several observatories to gain planetary reach. The eight radioastronomical stations currently in the network include Iram in Spain, the Large Millemeter Telescope (LMT) in Mexico, the Submillimeter Telescope (SMT) in Arizona, the James Clark Maxwell Telescope (JCMT) and the Submillimeter Array (SMA) in Hawaii, the South Pole Telescope (SPT) in Antarctica, the Atacama Large

Millimeter Array (Alma) and the Atacama Pathfinder Experiment (Apex) in Chile. Each of these instruments is located at altitude to reduce the atmospheric absorption of the signals. The whole set creates a virtual observatory with a 5,000-km aperture. Other observatories -the Greenland Telescope Project, in Greenland, and the Iram Noema interferometer, on the Bure plateau in the French Alps - will extend the network further and improve its performances.

**A Long Wait**

Once the concept was put forward, it then just had to be achieved, despite a multitude of observational and technical constraints. For example, weather conditions need to "cooperate" so that the VLBI network can avail of crystal skies simultaneously at eight places on four continents, as observations are not possible in the rain. The 1.3 mm wavelength at which the signals are detected is absorbed and emitted by water. Another constraint is imposed by the use of the Alma telescope in Chile, the most requested radio observatory in the world. Ultimately, the EHT teams only had a two-week window each year in which to attempt the group observations. They had to wait until April 2017 to have four full, clear nights, two for SgrA* and two for M87*. However, there was no question of seeing an image directly on a screen! Building a high-resolution image by VLBI requires the combination of the signals captured by the various network aerials. To do so, atomic clocks are used to measure the arrival time of the signals to one tenth of a billionth of a second, compare them in real time and triangulate with their point of origin to reconstitute an overall image. With eight observatories spread around the globe, including in places with poor internet links, the EHT scientists had to record the data separately and store them on hard disks to combine them subsequently.

The mass of data collected exceeded anything that had ever been done in all scientific fields together: one night of observation collected 2 petabytes of data, as much as in one complete year of experiments at the LHC, the CERN large-hadron collider that led to the discovery of the Higgs-Englert boson in 2012, following analysis of 4 million billion proton-proton collisions...

The hard disks of data stored in Antarctica then had to wait until December and the end of the long glacial winter, to be transported in secure flight conditions to join the several thousand hard disks centralised at the MIT Haystack Observatory in Massachusetts and the Max Planck Institute for Radio Astronomy in Bonn. There, clusters of supercomputers started to process the mountain of data, a task that will take several months to obtain just a few pixels of image of the massive black holes at the centre of the Milky Way and the furthest galaxy, M87.

The results will probably not be published until later in 2018. But, as suggested by the digital simulations recently done by the EHT scientists [15], the recomposed image obtained should resemble a brilliant crescent surrounding a black disk, placed on the side where the hot spot of the accretion disk is moving towards the observer.

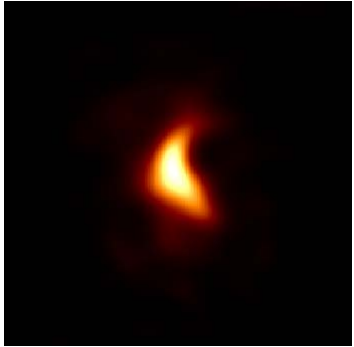

**Figure 8.** Synthetic image of the time variable black hole SgrA* that could be recovered with an array of 8 radiotelescopes and averaging 8 epochs(from [16]).

As could already be predicted from our simulations forty years ago (figure 1), by squinting your eyes to reduce the ocular resolution, the clear outline of a black hole surrounded by its accretion disk is indeed the black silhouette of its event horizon surrounded by the brilliant spot of the disk amplified by the Doppler effect! A very different, although improbable result, would mean that general relativity is incorrect in very strong gravitational fields, and new physics would be necessary.

**The golden age of relativistic astrophysics**

As well as the hope of capturing the first "photos" of black holes, EHT astrophysicists hope to garner a lot of information that will enable them to better understand the very special physics that reign in the environment close to black holes, namely the gigantic jets of particles and radiation that some project into space at speeds close to that of the speed of light [16]. This is the case for M87*, where the jets are larger than the galaxy itself. If SgrA* produces jets, they are too small or not bright enough to be detected until now. The jets play an important role in the evolution of galaxies; for example, by heating interstellar space, they can prevent cooling of the gas that allows stars to form. The most probable physical explanation is that they are produced by the twisted magnetic fields associated with black holes, fed in energy either by an accretion disk, or by conversion of some of the rotational energy of the black holes themselves. EHT telescopes are able to record the polarisation of the radiation emitted by the accretion disk of the galactic black hole SgrA* and to draw up the map of the magnetic field near its event horizon, which would perhaps reveal the physical mechanisms at the origin of the jets. VLBI observations in 2015 started to provide some clues as to the structure of SgrA*'s magnetic field, hinting at the hypothesis of a rapidly rotating black hole [17].

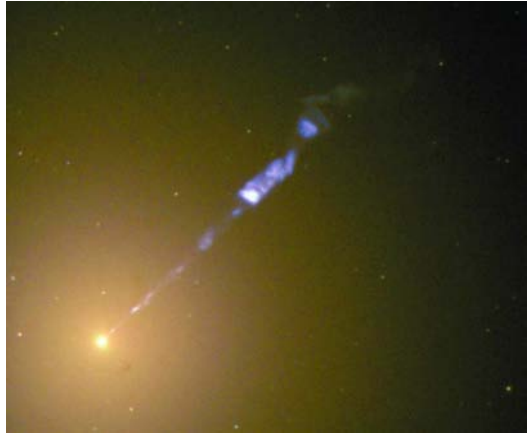

**Figure 9.** The radio jet of the elliptical galaxy M87 is also visible in the optical range. Detailed observations with the Hubble Space Telescope show various substructures, perfectly aligned from the central nucleus up to distances much greater than the size of the galaxy itself (© STSci/NASA).

An accurate description of black holes and their immediate environment still promises even more spectacular developments in the near future, thanks to a set of very different instruments and methods. The optical interferometer Gravity [18], being built at the Very Large Telescope (VLT) at the European Southern Observatory in Chile, and the next generation of optical telescopes in the 30 metre diameter class, will be able to follow the stars around Sgr A* orbiting at only a few hundred times the radius of the black hole, and measure the precession of their pericentres to deduce the angular momentum (spin) of the black hole. The radio interferometer SKA (Square Kilometer Array) [19], built in South Africa and Australia, will be able to follow the orbits of pulsars around the galactic black hole, timing them ultra-precisely to test their properties. The Lisa spatial interferometer (Laser Interferometer Space Antenna) [20], once in orbit, will be used to capture the gravitational waves emitted when small compact objects turn around supermassive black holes in nearby galaxies.

Relativistic astrophysics, still in its infancy in the 1970s due to a lack of experimental resources, is now entering a golden age. Already with the EHT, by capturing the signals of what is happening immediately close to a supermassive black hole, astrophysicists will be able to test Einstein's theory of general relativity in the most extreme conditions. These measurements will complete historical detections of gravitational waves from 2015 [21], produced when pairs of stellar-mass black holes collide, providing the best evidence so far of the existence of black holes. EHT data will be able to give us the final proof.